\documentclass[preprintnumbers, prd, twocolumn, showpacs,floatfix,preprintnumbers,
superscriptaddress, nofootinbib]{revtex4}
\usepackage{graphicx}
\usepackage{epsfig}
\usepackage{bm}
\usepackage{amssymb}
\usepackage{float}
\usepackage{amsmath}
\usepackage{subfigure}
\usepackage{dcolumn}
\usepackage{cancel}
\usepackage[colorlinks]{hyperref}
\usepackage[usenames,dvipsnames]{color}
\hypersetup{
     breaklinks=true,
    pdfstartview={FitH},    
    colorlinks=true,       
    linkcolor=blue,          
    citecolor=red,        
    filecolor=magenta,      
    urlcolor=blue,           
    anchorcolor=green,      
    linktocpage=true
}

\begin{document}
\title{Growth of Perturbations using Lambert$W$ Equation of State}

\author{Manisha Banerjee}
\email{banerjee.manisha717@gmail.com}
\affiliation{Department of Physics, Visva-Bharati, Santiniketan -731235, India}
\author{Sudipta Das\footnote{Corresponding author}}
\email{sudipta.das@visva-bharati.ac.in}
\affiliation{Department of Physics, Visva-Bharati, Santiniketan -731235, India}

\author{Abdulla Al Mamon}
\email{abdulla.physics@gmail.com}
\affiliation{Department of Physics, Vivekananda Satavarshiki Mahavidyalaya (affiliated to the Vidyasagar University),Manikpara-721513, West Bengal, India}

\author{Subhajit Saha}
\email{subhajit1729@gmail.com}
\affiliation{Department of Mathematics,
Panihati Mahavidyalaya, Kolkata 700110, West Bengal, India}

\author{Kazuharu Bamba}
\email{bamba@sss.fukushima-u.ac.jp}
\affiliation{Division of Human Support System, Faculty of Symbiotic Systems Science, Fukushima University, Fukushima 960-1296, Japan}

\pagestyle{myheadings}

\begin{abstract}
Recently, a novel equation of state (EoS) parameter for dark energy has been introduced which deals with a special mathematical function, known as the Lambert$W$ function. In this paper, we study the effect on the growth of perturbations for the Lambert$W$ dark energy model. We perform the analysis for two different approaches. In the first case we consider the universe to be filled with two different fluid components, namely, the baryonic matter component and the Lambert$W$ dark energy component, while in the second case we consider that there is a single fluid component in the universe whose equation of state parameter is described by the Lambert$W$ function. We then compare the growth rates of Lambert$W$ model with that for a standard 
$Lambda$CDM model as well as the CPL model.  Our results indicate that the presence of Lambert$W$ dynamical dark energy sector changes the growth rate and affects the matter fluctuations in the universe to a great extent.
\end{abstract}
\maketitle
\section{Introduction}
The well-established fact that the Universe is at present undergoing an accelerated expansion has been reinforced by a number of observational evidences obtained from the Supernovae data, Cosmic
Microwave Background Radiation (CMBR) data, Baryon Acoustic Oscillations (BAO) data, Large Scale Structure (LSS) of spacetime and many more. This led to the consideration of a replusive gravity component called Dark Energy (DE) which is usually characterized by an dynamical effective equation of state $w_{eff}$. A large number of functional forms for the equation of state parameter have been studied to account for this unknown component. For reviews on the various Dark Energy candidates, one can refer to \cite{tsujikawareview, matarresereview, Friemannreview, Tsujikawabook} . The proposed candidates for the equation of state parameter for Dark Energy are constrained with observational data in order to check the viability of a particular model. As nothing much is known about the mysterious Dark Energy component, search is still on to find a suitable candidate for Dark Energy. In this context, recently an {\it equation of state} (EoS) parameter has been proposed \cite{sahaw, mamon2005} which deals with a special mathematical function, known as the Lambert$W$ function. With this proposed form of EoS parameter, the  evolution of the Universe has been studied. It has been found that the Lambert$W$ EoS parameter for Dark Energy can successfully explain the evolutionary history of the Universe starting from an early acceleration phase and passing through a deceleration phase before entering into a late-time acceleration phase \cite{sahaw}. The functional form of the proposed Lambert$W$ EoS parameter is not very simple and straight forward. But the advantage is that a single EoS for Dark Energy can explain both the early and late-time acceleration of the universe at one go and thus is worth studying. 
\par In the present work we perform the perturbative analysis for the Dark Energy model described by Lambert$W$ equation of state parameter in order to have a better understanding of the effect of this particular Dark Energy model on the growth of perturbations. Because of the characteristic difference in the evolution of different dynamical DE models, they will have different impact in the matter power spectrum and accordingly will effect the structure formation differently. The study of perturbative effects for a particular dynamical DE model may be useful in discriminating between a cosmological constant model and various other models of dynamical dark energy. Future data from various ongoing as well as upcoming surveys such as eBOSS \cite{eBOSS}, DESI \cite{DESI}, LSST \cite{LSST} etc are expected to provide the cosmologists tools to perform precise measurements of the growth of structures in the universe. This will in turn allow us to identify viable dynamical dark energy models and their effectiveness as compared to a $\Lambda$CDM model.

\par In the following sections we provide a detailed description of the Lambert$W$ EoS and its cosmological implications. In the final sections, the results obtained from perturbative analysis are summarized. 

\section{Background cosmological scenario with the Lambert$W$ equation of state}
In mathematics, the Lambert$W$ function is defined as the multivalued inverse of the function $xe^x$ \cite{sahaw,mamon2005,Lambert1,Lambert2,Corless1,Corless2}, i.e.,
\begin{equation}\label{eq-defw}
   {\rm Lambert}W(y) \cdot e^{{\rm Lambert}W(y)}=y
    \end{equation}
    The Lambert$W$ function is also called ``product logarithm'' or the ``omega function''. Equation (\ref{eq-defw}) has two real solutions if $-\frac{1}{e}\le y<0$, and thus there are 
two real branches of the Lambert$W$ function \cite{branch}. Lambert's earlier work \cite{Lambert1,Lambert2} on the transcendental equation of the form
\begin{equation}
    x^m - x^n = (m-n)\nu x^{m+n}, ~~~m, n, \nu ~\mbox{are constants}
\end{equation}
led Euler \cite{Euler} to study the applicability of Eq. (\ref{eq-defw}). The $n^{th}$ derivatives of the Lambert$W$\footnote{Henceforth we shall simply write $W$ instead of Lambert$W$ in equations for our convenience.} function can be calculated as \cite{sahaw,mamon2005,hayes}  
    \begin{equation}\label{eq-deriw}
 W^n(y)=\frac{W^{n-1}(y)}{y^n[(1+W(y)]^{2n-1}}\phi^n_{k=1}\delta_{kn}W^k(y),  y \neq -\frac{1}{e}
\end{equation}
where $\delta_{kn}$ is the number triangle given by 
\[ \begin{array}{ccccc}
\phantom{+}1 & & & &\\
-2 & -1 & & & \\
\phantom{+}9 & \phantom{+}8 & \phantom{+}2 & & \\
-64 & -79 & -36 & -6 & \\
\phantom{+}625 & \phantom{+}974 & \phantom{+}622 & \phantom{+}192 & \phantom{+}24  
\end{array} .\]
Now, the first order derivative of $W(y)$ can be obtained from equation (\ref{eq-deriw}) as
\begin{eqnarray}\label{wprime}
W^{'}(y)&=&\frac{W(y)}{y[(1+W(y)]},  ~~{\mathrm if}~~   y \neq 0; \nonumber \\
&=&\frac{e^{-W(y)}}{[(1+W(y)]}
\end{eqnarray}
Thus, the antiderivative of $W(y)$ is obtained as
\begin{equation}
\int W(y)dy=y[W(y)-1+\frac{1}{W(y)}]+C
\end{equation}
in which $C$ is the arbitrary constant of integration. Additional interesting mathematical features of this special function can be found in \cite{branch,addf1}. It deserves mention here that many real-life applications of this function can be found in several branches of Physics, Mathematics, Computer Science and Engineering (for details, one might look into \cite{sahaw,Corless1,gr1,gr2} and the references therein).The motivation behind introducing the Lambert$W$ function in the cosmological context has been two-fold: 
\begin{enumerate}
    \item The Lambert$W$ function appears in solutions of those mathematical equations where the variable occurs both as a base as well as an exponent. To demonstrate, one may consider the solutions of the equations $e^x = 3x-5$ and $\text{ln}(4x)=x$. A vast archive of observational data has helped to establish that the behaviour of the Universe during the early inflationary phase was essentially exponential and the present phase of cosmic acceleration also shows similar behaviour. These exponential variations might suggest some interconnection with the Lambert$W$ function.
    \item This special mathematical function has been obtained implicitly while finding solutions of the continuity equation in the gravitationally induced adiabatic particle creation model \cite{chak1}. Thus, one cannot help but wonder if the Lambert$W$ function is somehow entangled with the cosmic substratum.
\end{enumerate}
\par Recently a novel EoS has been proposed which incorporates the Lambert$W$ function in a
special way \cite{sahaw, mamon2005}. Using the latest Hubble parameter dataset, Mamon and Saha \cite{mamon2005} have shown that this unified EoS, $w_{eff}$, can give rise to a late time accelerated phase of the universe preceded by a decelerated phase of expansion. They have also found that the present value of $w_{eff}$ is very close to the cosmological constant and, therefore, this new EoS might be considered as a possible unification of dark matter and dark energy. Furthermore, they have found that for the best-fit case, the differences between the Lambert$W$ model and the $\Lambda$CDM model are negligible around redshift $z\sim 0.67$. These deductions show that the cosmological implications of this Lambert$W$ EoS might be far-reaching. Motivated by the above facts, in the present work, we study the perturbative analysis for the dark energy model described by the Lambert$W$ EoS parameter in order to have a deeper understanding of the effect of this newly proposed cosmic fluid on the growth of perturbations.\\

\par We start with a spatially flat, homogeneous and isotropic Friedmann-Robertson-Walker universe described by the metric (we assume c = 1)
\begin{equation}
  ds^2 = dt^2 - a^2(t) [dr^2 + r^2 d\theta^2 + r^2 sin^2\theta d\phi^2]
\end{equation}
The background Einstein equations are obtained as 
\begin{equation} 
3 H^2 = 8 \pi G (\rho_m +\rho_{DE}) = 8 \pi G \rho
\end{equation}
\begin{equation}
\dot{H} + H^2 = - \frac{4\pi G}{3}(\rho + 3p)
\end{equation}
The conservation of the energy-momentum tensor will give the continuity equation in the form
\begin{equation} \label{cont}
    \dot{\rho}+3H(\rho+p)=0.
\end{equation}
As proposed by the authors in \cite{mamon2005}, the effective EoS parameter for the fluid is considered as 
\begin{equation}\label{lambertw}
w_{eff} = 
\left[\vartheta_1 \ln \left\{W\left(\frac{a}{a_0}\right)\right\} + \vartheta_2 \left\{W\left(\frac{a}{a_0}\right)\right\}^3
\right]
\end{equation}
where $a_0$ represents the present value of the scale factor of the universe and $\vartheta_1$, $\vartheta_2$
are dimensionless model parameters. The best fit values of $\vartheta_1$ and  $\vartheta_2$ have been found to be $-0.166 \pm 0.104$ and $-4.746 \pm 0.479$ within $1\sigma$ confidence limit \cite{mamon2005}. 
\par For the sake of completeness, we also provide the expressions for the energy density $\rho$ and the deceleration parameter $q$ obtained in \cite{sahaw} as
\begin{equation}
\begin{split}
\rho = \rho_{0}~{\rm exp}\Bigl[-3\Bigl\{\ln\left[W(a)\right][\vartheta_1 W(a) + \vartheta_1 + 1]\\ + W(a)\left(1 - \vartheta_1\right) + \frac{\vartheta_2}{12}{W(a)}^3\left[4 + 3 W(a)\right]\Bigr\}\Bigr]
\end{split}
\end{equation}
\begin{equation}
q=\frac{3}{2}\left[1 + \vartheta_1\ln \left[W(a)\right] + \vartheta_2 {W(a)}^3\right]- 1
\end{equation}
\section{Growth of Perturbations}
In this section we study the effect on the growth of perturbations for the Dark Energy model whose EoS is given by the Lambert$W$ function. We have performed the analysis for two different aspects : In the first case we consider the universe to be filled with two different fluid components, viz, the baryonic matter component (denoted by subscript $m$) and the Dark Energy component (denoted by subscript $DE$). We name this as the two fluid model in which the DE component is considered to be described by an equation of state parameter given by the Lambert$W$ function. We proceed by solving the perturbation equations for the two components of the universe. The details of analysis and the results are provided in the next section.   
\par
\begin{figure}
\begin{center}
\includegraphics[width=0.45\textwidth]{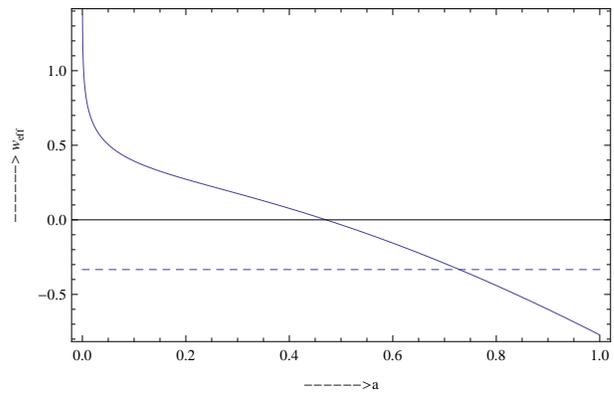}
\caption{\em Plot of $w_{eff}$ vs. $a$ for best fit values of $\vartheta_{1}=-0.166$ and $\vartheta_{1}=-4.746$}
\label{fig1}
\end{center}
\end{figure}

 In the second approach, we consider that there is a single fluid component in the universe whose equation of state parameter is depicted by the Lambert$W$ function. As mentioned by Saha and Bamba \cite{sahaw}, the effective equation of state $w_{eff}$, expressed in terms of Lambert$W$ function can provide a decelerating universe in the recent past followed by an accelerated expansion phase. This is also evident from figure \ref{fig1} which shows the variation of $w_{eff}$ with $a$ for the best fit values  $\vartheta_{1} =-0.166$ and $\vartheta_{2}=-4.746$ obtained in \cite{mamon2005}. As evident from the figure, $w_{eff}$ enters the regime $w_{eff} < -\frac{1}{3}$ (shown by dotted line in figure \ref{fig1}) in the recent past and prior to that was undergoing a decelerated expansion phase. Considering this variation of the Lambert$W$ function, we also assume that the universe is comprised of a single fluid which depicts the evolution history for matter dominated universe in the far past and then enters the accelerated regime in the near past. The perturbation equations are solved for such a single fluid model. 
\par The results obtained by following these two approaches are provided in the following subsections.   
\subsection{\bf Perturbations in a two fluid model }
In  this section, we consider that the present universe has two different components, namely, the normal baryonic matter whose energy density is denoted by $\rho_m$ and the Dark Energy (DE) component, whose energy density is denoted by $\rho_{DE}$. The equation of state parameter for the DE component is given by the Lambert$W$ function as given in equation (\ref{lambertw}). As the presence of DE component has an repulsive gravity effect, it is expected to affect the growth of matter over-density. We are interested to know how a Dark Energy sector, whose equation of state parameter is given by the Lambert$W$ function, effects the structure formation of the universe. We begin by considering the linearized Einstein's equations \cite{RAA2018, Jaber2017}
\begin{equation}\label{diff1}
\begin{split}
    a^2{{\delta}''_m}(a)+a\frac{3}{2}[1-w(a)\Omega_{DE}(a)]{\delta}'_m(a)-\frac{3}{2}[\Omega_m(a)\delta_m(a)\\+\Omega_{DE}(a)\delta_{DE}(a)]=0
\end{split}
\end{equation}
\begin{equation}\label{diff2}
\begin{split}
    a^2{\delta}''_{DE}(a)+a\frac{3}{2}[1-w(a)\Omega_{DE}(a)]{\delta}'_{DE}(a)+\Bigl(\frac{c_s^2k^2}{a^2H^2(a)}-\\\frac{3}{2}\Omega_{DE}(a)\Bigr)\delta_{DE}(a)-\frac{3}{2}\Omega_m(a)\delta_m(a)=0
\end{split}
\end{equation}
where $\delta_{DE} \equiv \frac{\delta\rho_{DE}}{\rho_{DE}} $ and $\delta_m \equiv \frac{\delta\rho_m}{\rho_m} $ represents the matter and DE density contrast respectively. Here a prime represents derivative with respect to $a$ and $k$ is the Fourier wave number. $w(a)$ is the equation of state parameter for the Dark Energy component (expressed as $w_{eff}$ in equation (\ref{lambertw})) and is given by the Lambert$W$ function.
\newline $c_s^2$ represents as usual the speed of sound for the Dark Energy component and is given by 
\begin{equation}\label{cs2}
c_s^2 \equiv \frac{\delta p}{\delta \rho} \equiv c_{ad}^2 +c_{eff}^2
\end{equation}
where $c_{ad}^2$ which is the adiabatic part of the sound speed and is given by  
\begin{equation}\label{cadiabatic}
c_{ad}^2 \equiv \frac{dp}{d \rho} = w(a)-\frac{a w' (a)}{3(1+w(a))} 
\end{equation} 
From Equation (\ref{lambertw}), we have the expression of $w'(a)$ as
\begin{equation}
w'(a)= \frac{v_1+3v_2{W(a)}^3}{a(1+W(a))}
\end{equation} 
which can be used to obtain the expression for $c_{ad}^2$ for the Lambert$W$ Dark Energy component. Here $ c^2_{eff}$ represents the non-adiabatic part of the speed of sound which has been chosen to be equal to $1$. 

\begin{figure}
\begin{center}
  \includegraphics[width=6cm, height=4cm]{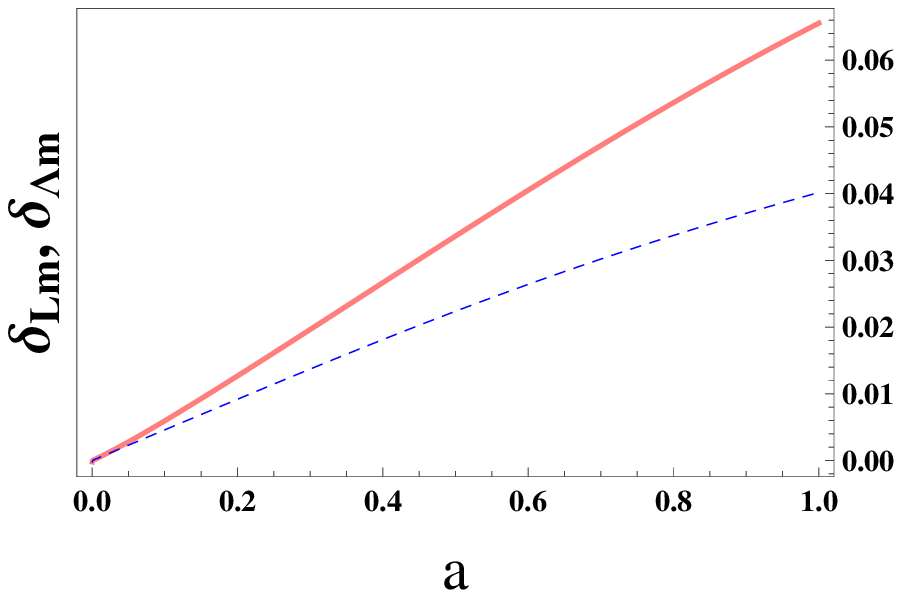}
  \hspace{1cm}\includegraphics[width=6cm, height=4cm]{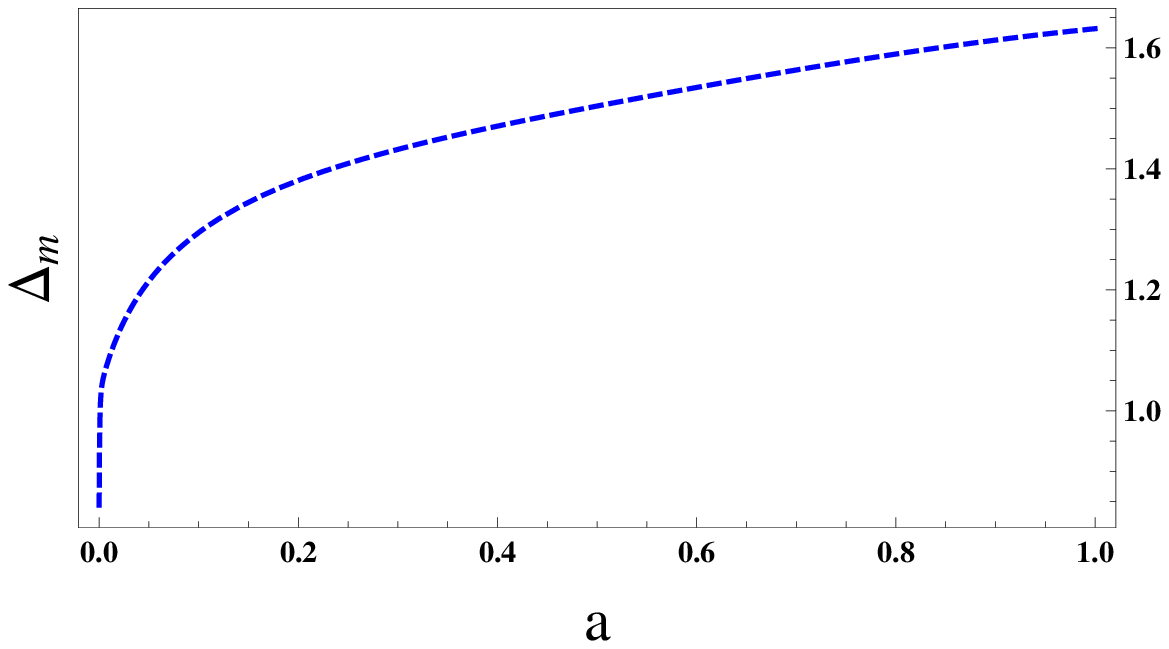}
\caption{\em The top panel shows the matter density contrasts $\delta_m(a)$ for two fluid model whose EoS is given by Lambert$W$ function (orange solid line) in comparison to the matter density contrast for a $\Lambda$CDM Dark Energy model (blue dashed line). The bottom panel shows the ratio of the density contrasts given by $\Delta_m=\frac{\delta_{Lm}}{\delta_{\Lambda m}}$}
\label{LL}
\end{center}
\end{figure}
We have obtained the matter density contrast $\delta_m(a)$ by solving the coupled differential equations (\ref{diff1}) and (\ref{diff2}) numerically for $c^2_{eff} = 1$. As already mentioned, we have considered that the Dark Energy component is characterised by the Lambert$W$ EoS parameter. The results have been displayed in figure \ref{LL}. The solid line in the top panel of figure \ref{LL} represents the matter density contrast for a two fluid model, one of which being the normal baryonic matter and the other one being the Dark Energy component whose EoS is given by the Lambert$W$ function. For comparison, we have also plotted the matter density contrast profile for a $\Lambda$CDM Dark Energy model, shown by the dashed blue line in the top panel of figure \ref{LL}. These results have been obtained for $c^2_{eff} =1$ and $c^2_{ad}$ given by equation (\ref{cadiabatic}). The bottom panel of figure \ref{LL} shows the ratio of the two density contrasts given by $\Delta_m=\frac{\delta_{Lm}}{\delta_{\Lambda m}}$, which provides a measure of the effect of the Lambert$W$ DE model on the growth of structure formation as compared to a $\Lambda$CDM model. It indicates that because of the dynamical nature of the Lambert$W$ EoS parameter, this particular form of DE component will have more effect on the structure formation of the universe. 

\par For the entire analysis, we have set our initial conditions assuming that at the beginning the contribution due to the DE component was very small and the modes were well inside the Hubble horizon. We have chosen $\delta_{m}(a_{\mathrm{ini}}) = 10^{-5}$ and $k=0.01 Mpc^{-1}$. For the DE sector, the initial contribution has been set at  $\delta_{DE}(a_{\mathrm{ini}})=10^{-8}$ which is almost negligible.
\begin{figure}[ht]
\begin{center}
\includegraphics[width=6cm, height=4cm]{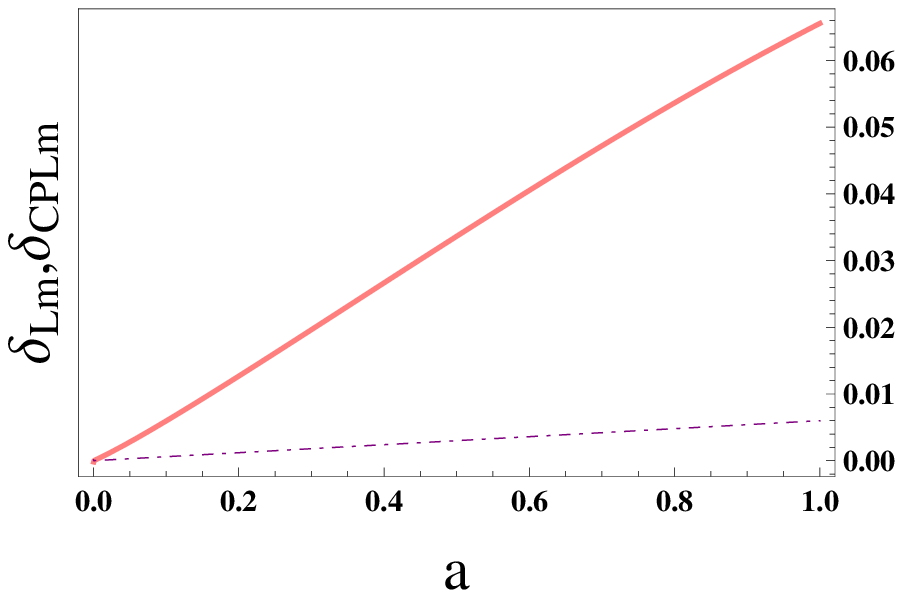}\hspace{1cm}
\includegraphics[width=6cm, height=4cm]{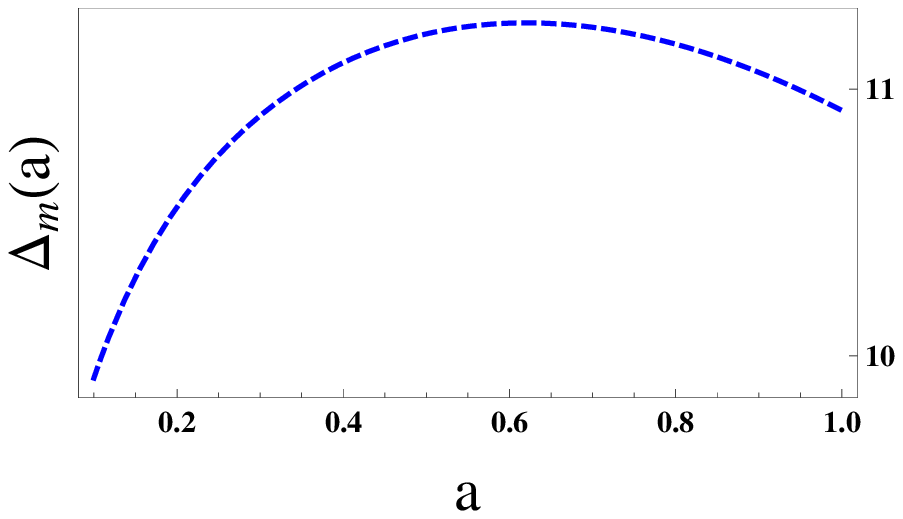}
\caption{\em The matter density contrasts $\delta_m(a)$ for Lambert$W$ Dark Energy model (pink solid line) and CPL Dark Energy model (purple dot-dashed line) (top panel) for $c^2_{ad}=0$. The bottom panel shows the ratio of the two density contrasts given by $\Delta_m=\frac{\delta_{Lm}}{\delta_{CPLm}}$}
\label{LCPL}
\end{center}
\end{figure}
 \par In figure \ref{LCPL}, we have done a similar analysis and have plotted the matter density contrasts $\delta_m(a)$ for two different Dark Energy Fluid sectors :  Lambert$W$ EoS parameter of DE and  the Chevallier-Polarski-Linder (CPL) parametrization of DE \cite{CPL1, CPL2} given by $w_{CPL}=w_0+w_1(1-a)$. The results have been displayed in figure \ref{LCPL}. For the analysis, we have considered $w_0=-1.17$ and $w_1=0.35$ \cite{Qi}. In this particular case,  we have performed the analysis for $c^2_{ad}=0$ as for CPL model the expression for $c^2_{ad}$ encounters an infinite value during the evolution and thus numerical integration can not be performed. We have chosen $c^2_{eff} =1$ as before. In figure \ref{LCPL} also, the pink solid line in the top panel represents the matter density contrast for the Lambert$W$ DE model and the  purple dashed line represents the same for CPL Dark Energy model. As before, the bottom panel shows the ratio between the matter density contrasts for these two sectors given by $\Delta_m=\frac{\delta_{Lm}}{\delta_{CPLm}}$ which shows clearly that the Lambert $W$ Dark Energy model has more significant effect on the growth of matter perturbations. 
 \\
\par
\begin{figure}[ht]
\begin{center}
\includegraphics[width=8cm, height=5cm]{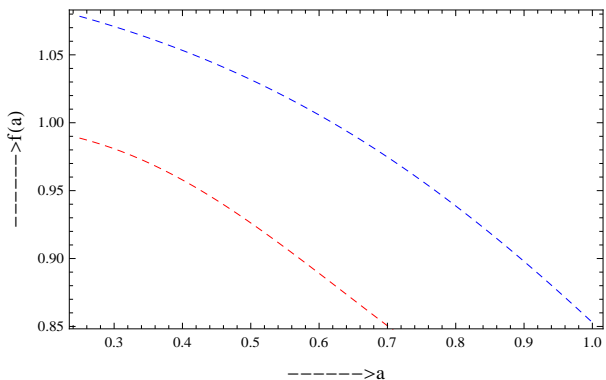}\hspace{1cm}
\caption{\em Logarithmic growth function  $f(a)=\frac{d~log \delta_m (a)}{d~ log a}$
for the Lambert$W$ Dark Energy model (blue dashed line) and $\Lambda$CDM model (red dot-dashed line) for $c^2_{ad}$ given by equation (\ref{cadiabatic}).}
\label{GFTF}
\end{center}
\end{figure}

 We have also evaluated numerically the logarithmic growth function, $f(a)=\frac{d~log \delta_m (a)}{d~ log a}$ for the two fluid model. In figure \ref{GFTF} we show the theoretical predictions for the logarithmic growth function for the Lambert$W$ Dark Energy model (shown by the blue dashed line) and compare it with the $\Lambda$CDM model (shown by red dot-dashed line). It is evident from the figure that the  growth of structure is slower in a $\Lambda$CDM model as compared to a Lambert$W$ model as we have obtained a slower logarithmic growth rate in case of  $\Lambda$CDM model. The reason for this being the form of the equation of state parameter $w(a)$ or equivalently $w_{eff}(a)$ given by equation (\ref{lambertw}). Because of the dynamical nature and stiffness of $w(a)$, it might have more visible effects on the growth of structures as compared to a $\Lambda$CDM model.  
Using the growth data from future surveys such as eBOSS, DESI, Euclid, or WFIRST, one should be  clearly able to identify whether the dynamical DE models are more preferable compared to $\Lambda$CDM models or not. 

\subsection{\bf Perturbations in a Single-fluid model}
In this subsection, we study the effect of perturbations in a Lambert$W$ equation of state parameter model from a different aspect. We assume that the universe has a single fluid component whose equation  of  state  parameter  is  given  by  the  Lambert$W$ function as given in equation (\ref{lambertw}).  As  mentioned  by  Saha  and Bamba \cite{sahaw}, the effective equation of state parameter $w_{eff}$ provides a decelerated expansion phase of the  universe ($w_{eff} > -\frac{1}{3}$) in the recent past followed by an accelerated expansion phase ($w_{eff} < -\frac{1}{3}$), which is evident from figure \ref{fig1}. This single fluid is thus capable of reproducing the evolution history of matter dominated as well as Dark Energy dominated phases of the universe at one go and thus is worth studying.
 
\par We begin by considering the perturbation equation for a single fluid model given by \cite{Tsujikawabook} 
\begin{equation}
\begin{split}
 \delta^{\dagger\dagger} + {\mathcal{H}}(1 + 3 {c_s}^2 -6 w_{eff}) \delta^{\dagger} \\-\left[\frac{3}{2}  {\mathcal{H}}^2 (1 - 6 {c_s}^2 -3 w_{eff}^2+8 w_{eff})-{c_s}^2 k^2\right]\delta = 0
\end{split}
    \end{equation}
    where a $\dagger$ sign indicates differentiation with respect to conformal time $\eta =\frac{dt}{a}$. Rewriting the equation by changing the argument from $\eta$ to $a$, we get
\begin{equation}\label{sf11}
\begin{split}
    a^2 H(a)\left[ \delta''(a) a^2 H(a) + 2 a H(a) \delta'(a) + a^2 H'(a) \delta'(a)\right] +\\ (1 + 3 {c_s}^2 -6 w_{eff}) a^3 H^2(a) \delta'(a) 
    -\\ \left[\frac{3}{2} a^2 H^2(a) (1 - 6 {c_s}^2-3 w_{eff}^2+8 w_{eff})-{c_s}^2 k^2\right]\delta(a) = 0
\end{split}    
\end{equation}
 where a prime indicates differentiation with respect to $a$. $w_{eff}$ is the effective equation of state parameter for the fluid component (here corresponds to the Lambert$W$ equation of state parameter given by equation (\ref{lambertw})) and ${c_s}^2$ is the sound speed for the fluid component given by equations (\ref{cs2}) and (\ref{cadiabatic}). $k$ as usual represents the Fourier wave number and has been chosen to be equal to $0.01 Mpc^{-1}$. We choose the same set of initial conditions ($\delta(a_{\mathrm{ini}}) = 10^{-5}$) for solving the perturbation equation. 
\par
\begin{figure}[ht]
\begin{center}
\includegraphics[width=7cm, height=4.5cm]{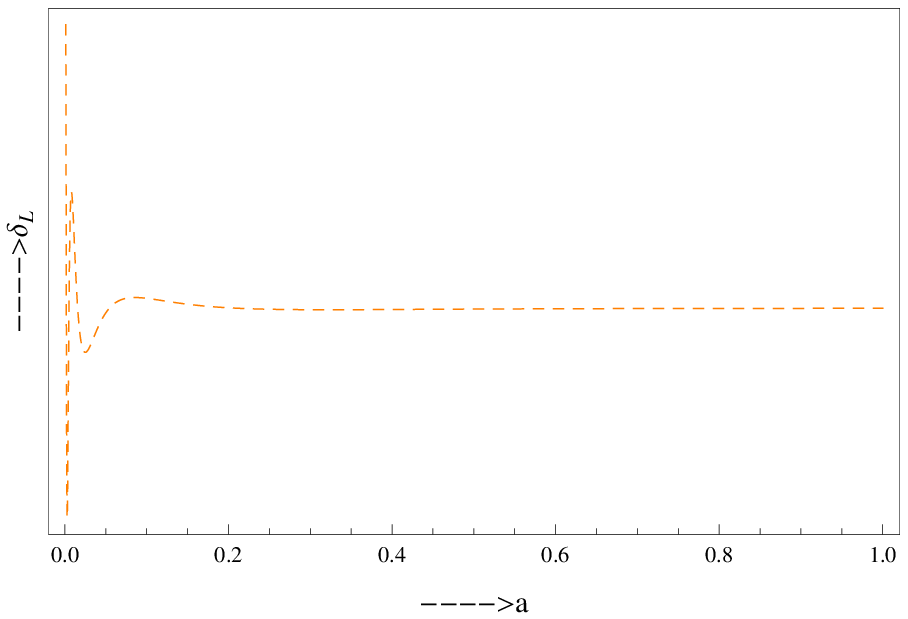}\hspace{1cm}\includegraphics[width=7cm, height=4.5cm]{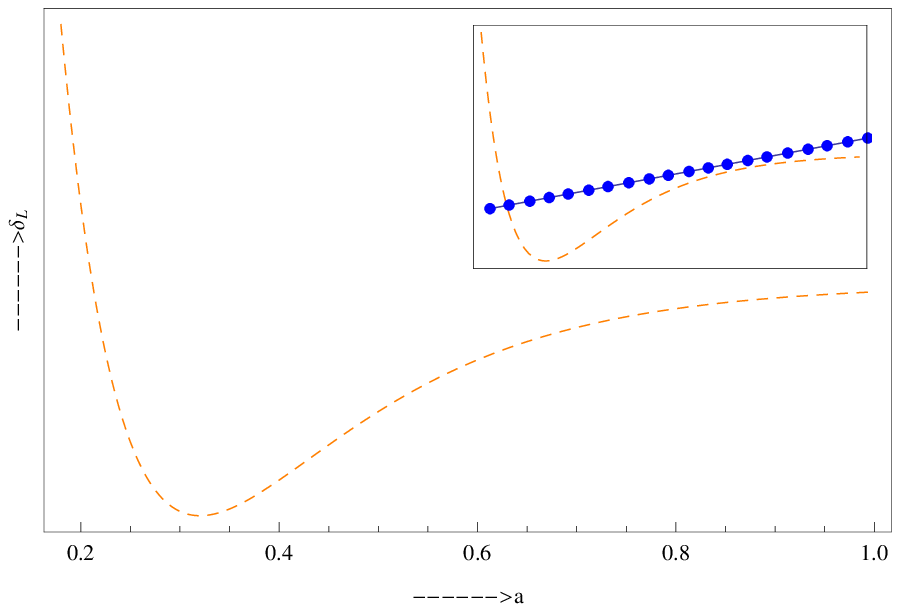}
\caption{\em Variation of density contrast $\delta_L(a)$ for the 
single fluid model. The bottom panel shows the zoomed view where the blue line in the offset represents the density contrast for the normal matter sector ($\delta_m \sim \frac{1}{a}$)}
\label{delta_single_fluid_fig}
\end{center}
\end{figure}

 Figure \ref{delta_single_fluid_fig} shows the variation of density contrast $\delta$ with scale factor $a$ for the single fluid model obtained by numerically solving equation (\ref{sf11}). 
The top panel of fig \ref{delta_single_fluid_fig} shows that for the single fluid model, there are some initial fluctuations at the very early stage and then there is steady growth in the matter density. The zoomed view of the density contrast, excluding the initial fluctuations, has been provided in the bottom panel of fig \ref{delta_single_fluid_fig} in order to have a clear picture of the steady growth rate. It is obvious from the figure that once the initial fluctuations settle down, the Lambert$W$ fluid depicts features similar to a normal matter sector which can give rise to the present observed structures of the universe. In the offset of the bottom panel of \ref{delta_single_fluid_fig}, the blue dotted line represents the $\delta_m \sim \frac{1}{a}$ curve as expected for a normal baryonic matter sector. As evident from the figure, the growth rate for $\delta_{L}$ trails the $\delta_m$ curve and thus reinforces the claim that the Lambert$W$ single fluid model reproduces the evolution history of matter dominated as well as Dark Energy dominated phases of the universe at one go.

\section{Discussion}
We have studied a DE model whose equation of state parameter is characterized by Lambert$W$ function given by equation (\ref{lambertw}). This model has been analyzed at the background level by Saha and Bamba \cite{sahaw} and the observational constraints on various model parameters have been studied by Mamon and Saha \cite{mamon2005}.  In this work, we extend the analysis by investigating how a Lambert$W$ equation of parameter for Dark Energy can affect the growth of structures in the universe. It has been observed that the presence of Lambert$W$ dynamical dark energy sector changes the growth rate and affects the matter 
fluctuations to a great extent. We have compared the growth rates of Lambert$W$ model with that for a $\Lambda$CDM model as well as CPL model and in both the cases it has been observed that the growth of matter fluctuations is more in Lambert$W$ DE model. The growth data from future DE surveys will allow us to decide whether such dynamical DE models are more preferable compared to $\Lambda$CDM models or CPL models or not.\\

\section{Acknowledgement}
SD acknowledges the financial support from SERB, DST, Government of India through the project EMR/2016/007162. SD would also like to acknowledge IUCAA, Pune for providing support through associateship programme. The work of KB was supported in part by the JSPS KAKENHI Grant Number JP 25800136 and Competitive Research Funds for Fukushima University Faculty (19RI017).

\end{document}